\documentclass[a4paper,11pt]{article}

\usepackage{jcappub}
\usepackage{amsfonts}
\usepackage{amsmath}
\usepackage{url}

\title{Evolution of the phase-space density and the Jeans scale for dark matter derived
from the Vlasov-Einstein equation}

\author[a]{O. F. Piattella, }\author[a]{D. C. Rodrigues, }\author[a]{J. C. Fabris, }
\affiliation[a]{Departamento de F\'isica, Universidade Federal do Esp\'irito Santo, Vit\'oria, ES, Brazil}
\author[b]{and J. A. de Freitas Pacheco}
\affiliation[b]{Universit\'e de Nice-Sophia Antipolis, Observatoire de la C\^ote d'Azur, Laboratoire Lagrange, Nice Cedex 4, France}

\emailAdd{oliver.piattella@pq.cnpq.br}
\emailAdd{davi.rodrigues@ufes.br}
\emailAdd{fabris@pq.cnpq.br}
\emailAdd{pacheco@oca.eu}

\abstract{We discuss solutions of Vlasov-Einstein equation for collisionless dark matter particles in the context of a flat Friedmann universe. We show that, after decoupling from the primordial plasma, the dark matter phase-space density indicator $Q = \rho/(\sigma_{\rm 1D}^2)^{3/2}$ remains constant during the expansion of the universe, prior to structure formation. This well known result is valid for non-relativistic particles and is not ``observer dependent'' as in solutions derived from the Vlasov-Poisson system. In the linear regime, the inclusion of velocity dispersion effects permits to define a physical Jeans length for collisionless matter as function of the primordial phase-space density indicator: $\lambda_J = (5\pi/G)^{1/2}Q^{-1/3}\rho_{\rm dm}^{-1/6}$. The comoving Jeans wavenumber at matter-radiation equality is smaller by a factor of 2-3 than the comoving wavenumber due to free-streaming, contributing to the cut-off of the density fluctuation power spectrum at the lowest scales. We discuss 
the physical differences between these two scales. For dark matter particles of mass equal to 200 GeV, the derived Jeans mass is $4.3\times 10^{-6}~M_\odot$.}

\keywords{Dark Matter, Velocity dispersion, Jeans mechanism, Phase space density, Vlasov-Einstein equations.}

\arxivnumber{1306.3578}

\begin{document}

\maketitle
\flushbottom

\section{Introduction}

Different observations suggest that dark matter (DM) contributes six times more than baryonic matter to the global energy budget of the universe. The absence of relic candidates among particles issued from the standard model (excepting neutrinos, which have an adequate abundance but are not massive enough and have additional difficulties related to their free-streaming length) led physicists to examine candidates issued from alternative models. In particular, Minimal Supersymmetric extensions of the Standard Model (MSSM) offer a plethora of candidates like gravitinos, photinos, s-neutrinos among others. Presently, the most plausible candidate is the neutralino, the lightest super-symmetric particle \cite{Bertone:2004pz}. However, there 
are some embarrassing problems: up to now, no signal of super-symmetry has been seen in experiments performed 
with the Large Hadron Collider (LHC) at CERN, since the decay of B-mesons does not indicate the presence of ``exotic" particles like charginos and/or neutralinos \cite{Aaij:2012ac,Bechtle:2012zk}. Moreover, results from direct search experiments are also under tension. In fact, a positive signal modulated with a period of about one year is claimed to be present in data from experiments like DAMA/LIBRA, CoGent and CRESS-II \cite{Bernabei:2010ke,Kelso:2011gd,Arina:2011zh,Hooper:2012ft} that are 
not confirmed by the more sensitive experiment XENON100 \cite{Aprile:2010um}. In addition, cosmological simulations 
predict too much DM at the centre of galaxies as well as a large number of satellites, in disagreement with 
observations \cite{Perivolaropoulos:2008ud,Donato:2009ab,DelPopolo:2010rj,deNaray:2010zw,Ogiya:2011ta}. 
\par
Despite these difficulties, DM theory predicts correctly the characteristics of large structures seen in the universe, the so-called ``cosmic web'' and, without such a component, it would be difficult to 
reach the non-linear regime of growth of primordial density fluctuations in an adequate time-scale necessary to 
explain the formation of galaxies \cite{Blumenthal:1984bp,Davis:1985rj,Navarro:1995iw}. 
\par
Assuming the reality of such component, we are still confronted with some problems like the relaxation 
processes leading structures to a state of near dynamical equilibrium. After decoupling from the primordial 
plasma, DM particles evolve as a collisionless system described by its phase-space distribution or its fine-grained distribution function $f(t, x_i, P_i)$, where $t$ is the time, $x_i$ the position and $P_i$ the momentum of the 
particle. In general, the fine-grained distribution is not very useful to follow the 
evolution of real physical systems. In practice, it is preferable to work with an average of $f$ over a 
finite volume of phase-space, i.e. with the coarse-grained distribution function $F(t,x_i,P_i)$. Such averaging process leads to $F \leq f$ \cite{tremaine1986,tremaine1987erratum}. An indicator of the coarse-grained phase-space density, commonly used in the literature \cite{Hogan:2000bv} reads
\begin{equation}\label{Qind}
 Q = \frac{\rho}{(\sigma_{\rm 1D}^2)^{3/2}}\;,
\end{equation}
where $\rho$ is the mass density, $\sigma^2_{\rm 1D}$ is the one-dimensional velocity
dispersion of the particles. In general, the
parameter $Q$ differs from $F$ by a numerical factor of the order of the 
unity (see, for instance, ref.~\cite{Shao:2012cg}). The phase-space density indicator $Q$ evolves as the 
coarse-grained distribution function $F$ or, in other words, according to the \textit{Mixing~Theorems} \cite{tremaine1986,tremaine1987erratum,binney1987galactic}. These theorems say that processes leading to the relaxation of collisionless systems, e.g. phase mixing and energy and angular momentum mixing generated by violent relaxation, result in a 
decrease of the coarse-grained phase-space density. For example, cosmological simulations 
indicate that $Q$, evaluated in the central region of halos, decreases as these objects evolve 
either as a consequence of a continuous accretion of mass or important merger episodes, processes which favour 
the violent relaxation mechanism \cite{Peirani:2005kw} and hence supporting theoretical expectations \cite{Vass:2008re}.

Considering now the primordial universe, for temperatures $T > m$, DM 
particles of mass $m$ are in thermal equilibrium with the cosmic plasma. When the condition $T \sim m$ is reached, detailed balance is destroyed and at $T_{\rm cd} \sim m/24$ the annihilation rate drops below the expansion rate, fixing the abundance of DM particles with respect to photons. The temperature $T_{\rm cd}$ characterizes the chemical decoupling of non-relativistic DM particles and afterwards the density of DM particles varies as $a^{-3}$, where $a$ is the scale factor. However, DM particles still remain thermally coupled with fermions (leptons and quarks) present in the cosmic plasma through scattering processes. Therefore, during this phase the velocity of DM particles is determined by the temperature of the relativistic fermions. The kinetic decoupling at the temperature $T_{\rm kd}$ depends on MSSM parameters 
and different estimates indicate $T_{\rm kd} \sim$ 25 MeV for DM particles with 
mass $m \sim$ 100 GeV \cite{Bringmann:2006mu}. Hence, the values of the 
density and of the velocity dispersion at $T_{\rm kd}$ fix the corresponding value of the primordial phase-space indicator $Q$.  

Using the Vlasov-Poisson system of equations in a expanding background, in ref.~\cite{Peirani:2008bu} the authors 
have shown that after kinetic decoupling, the phase-space indicator $Q$ associated to DM particles remains constant during 
the matter dominated (Einstein-de Sitter) expansionary 
phase of the universe. The invariance of $Q$ is broken once the linear regime ends and the different mixing 
mechanisms become operative. However there is a flaw in this approach. The general solution of the distribution 
function resulting from the Vlasov-Poisson system depends on space coordinates \cite{lee2010classical,rein1997nonlinear} and the invariance of $Q$ is guaranteed only for observers situated at the origin of the coordinate frame. To summarise, the value of $Q$ is expected to decrease both before kinetic decoupling and during the non-linear regime. In this paper we focus on the interval between these two epochs and we show, using the Vlasov-Einstein equation, that $Q$ remains constant at the background level.

In a second step, we derive the linearised Vlasov-Einstein equation and investigate the effects of a finite velocity dispersion of DM particles in the growth of primordial perturbations. There is a critical wavelength, equivalent to the Jeans length for a perfect fluid, below which density fluctuations do not grow, imposing a cut-off in the linear power spectrum. Such a critical length, although formally similar to the expression of the free-streaming damping length (see, for instance, refs.~\cite{Boyanovsky:2008he,Pavlov:2012zz} and references therein), has a different physical interpretation as we discuss in Sec.~\ref{Sec:Linthe}. In the present investigation, we assume that $Q$ remains constant
in the linear regime due to the absence of dissipative mechanisms for collisionless matter and, consequently, the
critical wavelength will depend on the primordial phase-space density. This
implies that the entropy per particle also remains constant according to eq.~\eqref{Eq:Sackur-Tetrode}.

The paper is organized as follows: in Sec.~\ref{Sec:Vlaseq} the Vlasov-Einstein equation is discussed; in Sec.~\ref{Sec:Linthe} the linear 
theory is revisited and in Sec.~\ref{Sec:Concl} the main conclusions are given. We use natural units $c = k_{\rm B} = \hbar = 1$ and metric signature $-,+,+,+$ throughout this paper.

\section{The Vlasov Equation}\label{Sec:Vlaseq}

Vlasov equation (VE), or collisionless Boltzmann equation (CBE), describes the evolution of collisionless
matter. Its generalization to an ensemble of freely moving particles in curved spaces can be found in 
several papers appeared in the literature in the past decades and comprehensive reviews can be found 
in ref.~\cite{Rendall:1996gx,bernstein2004kinetic,Andreasson:2005qy}. In particular, a study of the Vlasov-Einstein
system for spatially homogeneous spacetimes can be found in ref.~\cite{Okabe:2011nt}. In general, a four-dimensional 
manifold $\mathcal{M}$ characterized by a metric tensor $g_{ab}$ is considered, in which the 
metric is assumed to be time-orientable and the world-line of a particle with finite mass $m$ is a 
time-like curve. The unit future-directed tangent vector to this curve is the four-velocity of the 
particle $u^\mu$ and the four-momentum is $P^\mu = mu^\mu$. The possible values of the 
four-momentum form a hypersurface $P$ in the tangent bundle $T\mathcal{M}$ called the \textit{mass~shell} and 
the associated vector field $P^\mu$ is often called the ``geodesic spray". In this context, the 
distribution function $f$ represents the density of particles with given spacetime position and four-momentum. Moreover, geodesics 
have a natural lift to a curve on $P$, by taking its position and tangent vector together and hence 
defining a flow on $P$. The Vlasov equation expresses the conservation of $f$ along such a flow. 
\par
In our analysis, we found to be more convenient to use the modulus $p$ of the space components of the four-momentum
$P^i$ as an independent variable, i.e.
\begin{equation}
p^2 = g_{ij}P^iP^j\;,
\label{modulop}
\end{equation}
where $i,j = 1,2,3$. Since $g_{\mu\nu}P^\mu P^\nu = -m^2$, we define the energy 
as $E = \sqrt{-g_{00}}P^0$ and thus the modulus of the three-momentum satisfies the usual relation
\begin{equation}
E^2 = p^2 + m^2\;.
\label{energy}
\end{equation}
Thereby, using the variable $p$ or $E$ is only a matter of convenience. Considering a Friedmann-Lema\^itre-Robertson-Walker metric
\begin{equation}
ds^2 = -dt^2 + a(t)^2\gamma_{ij}dx^idx^j\;,
\label{metric1}
\end{equation}
where $\gamma_{ij}$ is the spatial comoving metric, we introduce the unit direction vector $\hat n^i$, satisfying the condition
\begin{equation}
\gamma_{ij}{\hat n^i}{\hat n^j}=1\;,
\label{unitvector}
\end{equation} 
so that the momentum can be written as
\begin{equation}\label{propmom}
 P^i = \frac{p}{a}\hat n^i\;.
\end{equation}
Therefore, VE can be cast in the following form:
\begin{equation}
\frac{\partial f}{\partial t} + \frac{dx^i}{dt}\frac{\partial f}{\partial x^i} + 
\frac{dp}{dt}\frac{\partial f}{\partial p} + \frac{d\hat n^i}{dt}\frac{\partial f}{\partial \hat n^i} = 0\;.
\label{vlasov1}
\end{equation}
We discuss solutions of this equation in the subsequent sections, where we assume a flat spatial metric $\gamma_{ij}$ and Cartesian coordinates so that $\gamma_{ij} = \delta_{ij}$.

\subsection{Zeroth-order solution and the parameter Q}

When the ``zeroth-order'' VE is considered in a Friedmann background, the second and the last terms 
in eq.~\eqref{vlasov1} can be neglected, due to the hypothesis of isotropy and homogeneity at the 
basis of metric eq.~\eqref{metric1}. Using the geodesic equations corresponding to the latter, eq.~\eqref{vlasov1} can be recast as
\begin{equation}
\frac{\partial f}{\partial t} - Hp\frac{\partial f}{\partial p} = 0\;,
\label{vlasov1-0}
\end{equation}
where we have introduced the Hubble parameter $H = \dot a/a$ and the dot denotes derivation 
with respect to the cosmic time $t$. The solution of eq.~\eqref{vlasov1-0} can be obtained from the 
method of characteristics, which implies in solving
\begin{equation}\label{characteristicseq}
 \frac{dp}{dt} = -Hp\;,
\end{equation}
i.e. the geodesic equation itself, whose solution $pa = \mbox{constant}$ expresses the decay of 
the momentum on an expanding background. Any function of the form $f = f(pa)$ is a solution 
of eq.~\eqref{vlasov1-0}. In particular, if DM particles are fermions, $f$ is the Fermi-Dirac 
distribution $f_{\rm FD}$. In this case, after decoupling, the phase-space density satisfies the 
solution $f_{\rm FD}=f_{\rm FD}(pa)$.
Hence, as already mentioned, the distribution function does not 
depend on space coordinates as it occurs when the Vlasov-Poisson system is considered \cite{peebles1993principles,Peirani:2008bu}.
\par
The particle number density $n$ is calculated as
\begin{equation}
n = \int d^3pf(pa) = \frac{4\pi}{a^3}\int^{\infty}_0f(x)x^2dx\;,
\label{density0}
\end{equation}
where $d^3p$ is the volume element in the proper momentum space ($p^i = aP^i$, see eq.~\eqref{propmom}) and 
we have introduced in the integrand the variable transformation $x = pa$. Notice that the particle 
number density scales as $n \propto a^{-3}$, irrespective of the particles being relativistic or not.

We introduce now the tensor $\omega^{ij} \equiv \langle v^iv^j\rangle$, where the proper velocity is 
defined by $v^i \equiv a dx^i/dt = p{\hat n^i}/E$, and the average has the following form:
\begin{equation}
\langle v^iv^j\rangle = \frac{1}{n}\int d^3pf(pa)\frac{p^2\hat n^i\hat n^j}{E^2}\;.
\label{tensoromega}
\end{equation}
Performing again the variable transformation $x = pa$, using eq.~\eqref{energy} and the solution
of the angular integral
\begin{equation}
\int d\Omega\;{\hat n^i}{\hat n^j} = \frac{4\pi}{3}\delta^{ij}\;,
\end{equation}
where $d\Omega$ is the angular part of $d^3p$, one obtains finally from eq.~\eqref{tensoromega}
\begin{equation}
\omega^{ij} = \frac{4\pi}{3na^3}\delta^{ij}\int^{\infty}_0 f(x)\frac{x^4}{x^2 + m^2a^2}dx\;.
\label{tensoromegab}
\end{equation}
In the shear-free case, the average (squared) one-dimensional velocity dispersion can be calculated from the relation 
$\sigma_{\rm 1D}^2 = \delta_{ij}\omega^{ij}/3$, which combined with eq.~\eqref{tensoromegab} gives
\begin{equation}\label{sigma1D}
\sigma^2_{\rm 1D} = \frac{4\pi}{3na^3}\int^{\infty}_0 f(x)\frac{x^4}{x^2 + m^2a^2}dx\;.
\end{equation}
When particles are non-relativistic (i.e. $m >> p$), we have $ma >> x$ and hence
\begin{equation}
\sigma_{\rm 1D}^2 \approx \frac{4\pi}{3m^2na^5}\int^{\infty}_0 f(x)x^4dx\;.
\label{vd}
\end{equation}
Since $n \propto a^{-3}$, eq.~\eqref{vd} indicates that $\sigma_{\rm 1D}^2 \propto a^{-2}$. From now on, we drop the subscript 1D.

\begin{figure}[htbp]
\centering
\includegraphics[width=0.7\columnwidth]{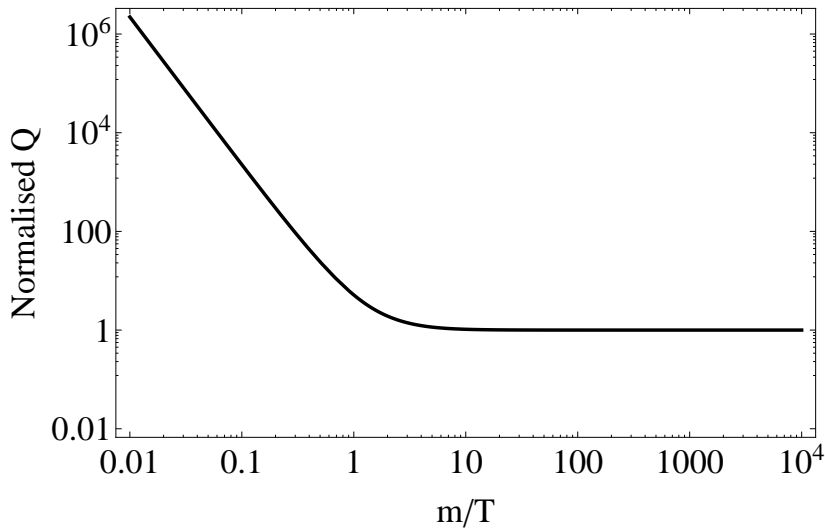}
\caption{Evolution of the phase-space parameter $Q$ normalised to the value of eq.~\eqref{parametro-Q}, as a function of $m/T$. We assumed $f(x) = 1/[\exp(\sqrt{x^2 + m^2a^2}) + 1]$ for numerical evaluation of the integrals defining the phase-space density indicator, cf. eqs.~\eqref{density0} and \eqref{sigma1D}. We have expressed the scale factor as $a \propto 1/T$.}
\label{fig0}
\end{figure}

Using eq.~\eqref{Qind} with $\rho = mn$ and eqs. \eqref{density0} and \eqref{vd}, one obtains for the phase-space indicator $Q$ 
\begin{equation}
Q = 4\pi\sqrt{27}m^4I_2^{5/2}I_4^{-3/2}\;,
\label{parametro-Q}
\end{equation}
where we have introduced the functional
\begin{equation}
I_n = \int_0^{\infty}f(x)x^ndx\;.
\end{equation}
Equation~\eqref{parametro-Q} shows that $Q$ does not depend on the scale factor $a$, remaining invariant during the expansion of the universe. $Q$ depends, as expected, on the 
fourth power of the particle mass, a property frequently used to put lower limits on the mass of DM
particles (see, for instance, ref.~\cite{Boyanovsky:2007ay}). From the definition of $Q$, using eqs.~\eqref{density0}, \eqref{sigma1D} and a Fermi-Dirac distribution, we plot in fig.~\ref{fig0} the evolution of $Q$ normalised to the value of eq.~\eqref{parametro-Q} as a function of $m/T$, where $T$ is the temperature, introduced via $a \propto 1/T$. Note that this normalised $Q$ equals unity approximately when $m/T \gtrsim 10$. This fixes a lower limit for the mass in order for our non-relativistic approximation be satisfied that is about few GeV if the kinetic decoupling temperature is about 25~MeV.

The primordial value of $Q$ defined at kinetic decoupling can be calculated in the following way (for the
sake of clarity, we recover all the physical constants). The relative abundance of DM particles with respect to photons is fixed at
chemical decoupling. Thus, if decoupling occurs when particles are non-relativistic, the DM density $\rho$ at chemical decoupling is given by
\begin{equation}
\rho(T_{\rm cd})=\frac{g}{\sqrt{8\pi^3}}\frac{m^{5/2}}{\hbar^3}\left(k_{\rm B}T_{\rm cd}\right)^{3/2}\exp\left(-\frac{mc^2}{kT_{\rm cd}}\right)\;,
\end{equation}
where $g$ is the degree of freedom of DM particles. From chemical decoupling up to kinetic decoupling, the DM density 
varies as $a^{-3}$. Thus, at kinetic decoupling the DM density will be
\begin{equation}
\rho(T_{\rm kd})=\rho(T_{\rm cd})\left(\frac{T_{\rm kd}}{T_{\rm cd}}\right)^3\;.
\end{equation}
The one-dimensional velocity dispersion of DM particles when thermal decoupling occurs is
\begin{equation}
\sigma^2=c^2\left(\frac{k_{\rm B}T_{\rm kd}}{mc^2}\right)\;.
\end{equation}
Combining these equations, one gets finally for the phase-space density indicator $Q$
\begin{equation}
Q=\frac{g}{\sqrt{8\pi^3}}\frac{m^4}{\hbar^3}\left(\frac{T_{\rm kd}}{T_{\rm cd}}\right)^{3/2}\exp\left(-\frac{mc^2}{k_{\rm B}T_{\rm cd}}\right)\;.
\end{equation}
For DM particles of mass $m\sim$ 100 GeV, the expected value of the primordial phase-space density indicator is 
$Q \sim 2\times 10^{16}~M_{\odot}$ pc$^{-3}$ km$^{-3}$ s$^{3}$, which should be compared with estimated values at the centre
of galaxies, i.e., $Q\sim 10^{-6}-10^{-7}~M_{\odot}$ pc$^{-3}$ km$^{-3}$ s$^{3}$ \cite{Dalcanton:2000hn}, indicating that during relaxation processes leading to structure formation, the value of $Q$ is reduced by 22-23 orders of magnitude.

Relations describing the evolution of the particle density or the velocity dispersion can be obtained in 
a differential form by taking the different momenta of eq.~\eqref{vlasov1-0}. In doing so, hereafter we 
assume the condition $\lim_{p \to \infty} p^n f(p) = 0$ for all $n$ (or at least for $n \le 7$). This 
condition is a natural one, in particular since the distribution function $f$ is expected to be of 
exponential form. Integrating eq.~\eqref{vlasov1-0} over the momentum space gives the conservation of the particle number density:
\begin{equation}
\frac{\partial n}{\partial t} + 3Hn = 0\;,
\label{continuity}
\end{equation}
which is essentially the differential form of eq.~\eqref{density0}. The first 
momentum of eq.~\eqref{vlasov1-0} is identically vanishing because of the isotropy of the spacetime. On the 
other hand, the second momentum gives
\begin{equation}
\frac{\partial}{\partial t}\int d^3p f \frac{p^2}{E^2}\hat n^i\hat n^j - H\int d^3p p\frac{\partial f}
{\partial p}\frac{p^2}{E^2}\hat n^i\hat n^j = 0\;.
\end{equation}
Introducing the definition of $\omega^{ij}$ [cf.~eq.~\eqref{tensoromega}], the above equation above can be rewritten as
\begin{equation}
\frac{\partial(n\omega^{ij})}{\partial t}- H\int d^3p p\frac{\partial f}
{\partial p}\frac{p^2}{E^2}\hat n^i\hat n^j = 0\;.
\label{sigma_a}
\end{equation}
Substituting $E^2 = p^2 + m^2$ into the integral above and integrating
by parts under the condition $\lim_{p \to \infty} \left[p^nf(p)\right] =0$, we get
\begin{equation}
H\int d^3p \left(5\frac{p^2}{m^2}-7\frac{p^4}{m^4}\right)f\hat n^i\hat n^j \approx 5Hn\omega^{ij}\;,
\end{equation}
where we have neglected contributions of order higher than $p^2/m^2$. Replacing the above result 
into eq.~\eqref{sigma_a}, and using also eq.~\eqref{continuity}, one obtains
\begin{equation}
\frac{\partial\omega^{ij}}{\partial t} + 2H\omega^{ij} = 0\;,
\end{equation}
and recalling that the velocity dispersion satisfies $\sigma^2 = \delta_{ij}\omega^{ij}/3$, we finally get
\begin{equation}
\frac{\partial \sigma^{2}}{\partial t} + 2H\sigma^{2} = 0\;. 
\end{equation}
This equation has a solution scaling as $\sigma^2 \propto a^{-2}$, consistent with the integral formula of eq.~\eqref{vd}.

It is interesting to verify whether the energy density evolution is affected by velocity dispersion 
effects. This can be done multiplying eq.~\eqref{vlasov1-0} by $E$ and integrating over the momentum space:
\begin{equation}
 \frac{\partial}{\partial t}\int d^3p Ef - H\int d^3p E p\frac{\partial f}{\partial p} = 0\;,
\label{energy1}
\end{equation}
where the first integral defines the energy density $\varepsilon$. It is easy to show that the two 
well-known solutions $\varepsilon = mn \propto a^{-3}$ for non-relativistic ($E \sim m$) particles 
and $\varepsilon = n\langle p\rangle \propto a^{-4}$ for relativistic ($E \sim p$) particles 
are reproduced from eq.~\eqref{energy1}. 

In the non-relativistic case, we consider now a first-order correction to the energy, i.e. $E \simeq
m + p^2/2m$, into eq.~\eqref{energy1}:
\begin{equation}
\label{energy2}
\frac{\partial\varepsilon}{\partial t} - H\int d\Omega\left[mp^3+\frac{p^5}{2m}\right]
\frac{\partial f}{\partial p}dp = 0\;.
\end{equation}
Now the energy density, using the first integral in eq.~\eqref{energy1} together with $E \simeq
m + p^2/2m$, is $\varepsilon = mn + mn\sigma^2/2$, i.e. it has the usual contribution from the 
rest mass plus an additional contribution due to a kinetic term. Performing an integration by 
parts in eq.~\eqref{energy2}, one finally obtains
\begin{equation}
\label{energy3}
\frac{\partial\varepsilon}{\partial t} + 3H\left(\varepsilon + \frac{1}{3}mn\sigma^2\right) = 0\;.
\end{equation}
In the energy conservation equation appears now an extra term formally comparable to an effective 
pressure $P = mn\sigma^2/3$, which is essentially the result obtained in \cite{Peirani:2008bu} via a 
different approach. Integration of eq.~\eqref{energy3} gives
\begin{equation}
\label{energy4}
\varepsilon = \frac{mn_0}{a^3} + \frac{1}{2}\frac{mn_0\sigma^2_0}{a^5}\;,
\end{equation}
where we have denoted as $n_0$ and $\sigma_0$ the present values (corresponding to $a_0 = 1$) of 
the particle number density and of the velocity dispersion, respectively. Notice that the solution 
of eq.~\eqref{energy3} is fully consistent with the adopted definition of the energy density, including 
a rest mass term and a kinetic contribution. This contribution is expected to be maximum when 
DM decouples from the primordial plasma (kinetic decoupling), representing a correction to the rest energy density 
of the order of $\sigma_{\rm kd}^2/2$, where $\sigma_{\rm kd}$ is the velocity dispersion at
kinetic decoupling. Since $\sigma_{\rm kd}^2/2 \sim (3/2)T_{\rm kd}/m$ 
and since typically $T_{\rm kd}\sim$ 25 MeV, this correction amounts to about $10^{-4}$ of the rest energy density, being
lately still smaller due to its fast decay with the scale parameter $a$.

\section{The linear theory revisited}\label{Sec:Linthe}

The growth of cold DM fluctuations in the linear regime is well studied and the theory is 
discussed in different papers and textbooks \cite{davis1977integration, Davis:1982gc, peebles1993principles, Kolb:1990vq, Dodelson:2003ft, Mukhanov:2005sc}, assuming in general an expanding background and considering the system constituted by the Vlasov and the Poisson equations. In particular, in refs.~\cite{davis1977integration,Davis:1982gc,peebles1993principles} the authors investigate the velocity dispersion for correlated pairs and triplets in the peculiar momentum space. In the present work we perform the analysis of the velocity dispersion for the full distribution function $f$ in a FLRW background, i.e. we address the Vlasov-Einstein equation and not the Vlasov-Poisson one.

The process of structure formation is independent of the DM nature since the growth of the density fluctuations
is controlled essentially by gravitation. However, at small scales the power spectrum of density fluctuations
is affected by the microphysics of DM particles. After kinetic decoupling, the main damping process is 
free streaming \cite{Hofmann:2001bi}. In the collisionless regime, DM particles move along geodesics in spacetime. If 
the proper distance travelled in a given time interval is larger than the proper wavelength of a perturbation, then 
such a structure will be smeared since particles propagate from an overdense to an underdense region.
In general, CDM is associated to a pressureless fluid and consequently all modes grow at the same rate. In fact, the 
velocity dispersion of DM particles, despite of being very small is not zero, permitting to define a critical length
below which perturbations do not grow, being a physical mechanism distinct from free streaming, as we will show later.   

The goal of this section is to review 
the linear theory including effects due to a finite velocity dispersion, under the Vlasov-Einstein 
formulation. As it was shown in 
the previous section, during the evolution of a collisionless system, the quantity $Q$, an 
indicator of the phase-space density, remains constant. In our analysis of the growth of the 
perturbations, we assume that $Q$ remains invariant also in the 
linear regime. This implies that variations of the velocity dispersion and of the density are not 
independent. The physical meaning of the constancy of $Q$ can be obtained if we consider 
the Sackur-Tetrode formula \cite{huang1987statistical} 
for the entropy per 
particle ($s = S/N$) 
of a Boltzmann gas:
\begin{equation}\label{Eq:Sackur-Tetrode}
s = \frac{3}{2}-\ln\left[n\left(\frac{6\pi\hbar^2}{m^2\sigma^2}\right)^{3/2}\right]= C - \ln Q\;,
\end{equation}
where $C$ is a constant. Thus, the entropy per particle is the logarithm of the inverse of 
the quantity $Q$. An evolution with constant $Q$ means that the entropy per particle remains constant. 
Using the Vlasov-Poisson system, in ref.~\cite{Pavlov:2012zz} analytical solutions for the linear equations have
been obtained in the case in which the velocity dispersion satisfies the condition $\sigma^2 \propto \rho^{2/3}$. This
condition, for an ideal gas, corresponds to an adiabatic expansion and corresponds essentially to the interpretation 
given above, since when $Q$ is constant the relation $\sigma^2 = (\rho/Q)^{2/3}$ is obtained.
\par
In our analysis, we consider a flat Friedmann spacetime with perturbations in the Newtonian or longitudinal gauge. Under these conditions
\begin{equation}
\label{metricperturbed}
ds^2 = -(1+2\Psi)dt^2 + a(t)^2\delta_{ij}(1+2\Phi)dx^idx^j\;,
\end{equation}
where the potentials $\Psi$ and $\Phi$ are functions of the cosmic time $t$ and of the comoving coordinates $x^i$.
\par
The distribution function $f$ (as other physical variables) can be split into ``zeroth'' and ``first'' order terms, i.e. as $f(t, x^i, P^i) = f^{(0)}(t, P^i) + f^{(1)}(t, x^i, P^i)$, or 
simply $f = f^{(0)}+f^{(1)}$. The integration of the distribution function $f$ over the proper momentum space gives
\begin{equation}
\int d^3pf=\int d^3pf^{(0)}+\int d^3pf^{(1)} \rightarrow n = n^{(0)}+n^{(1)}\;.
\label{density-2}
\end{equation}
 Replacing now the expression for $f$ into eq.~\eqref{vlasov1} and making use of eq.~\eqref{vlasov1-0} gives
\begin{equation}
\label{vlasov1-1}
\frac{\partial f^{(1)}}{\partial t}+\frac{p}{aE}\hat n^i\frac{\partial f^{(1)}}{\partial x^i}-
Hp\frac{\partial f^{(1)}}{\partial p}- \left(p\frac{\partial\Phi}{\partial t}+ \frac{E\hat n^i}{a}\frac{\partial\Psi}
{\partial x^i}\right)\frac{\partial f^{(0)}}{\partial p}=0\;,
\end{equation} 
where the term $dp/dt$ was computed from geodesics corresponding to the metric defined by eq.~\eqref{metricperturbed}.
\par
Before computing the momenta of eq.~\eqref{vlasov1-1}, it is useful to define the bulk velocity or the 
average proper velocity $V^i \equiv \langle v^i\rangle$, which in our case reads
\begin{equation}
V^i = \frac{1}{n}\int d^3pf\left(a\frac{dx^i}{dt}\right)\;,
\end{equation}
and it is identically vanishing in the background, due to isotropy. Since 
\begin{equation}
\frac{dx^i}{dt}=\frac{P^i}{P^0}=(1-\Phi+\Psi)\frac{p}{aE}\hat n^i\;,
\end{equation}
the equation for the bulk velocity can be recast as
\begin{equation}
\label{bulk0}
V^i=\frac{1}{(n^{(0)}+n^{(1)})}\int d^3p(1-\Phi+\Psi)\frac{p}{E}\hat n^i(f^{(0)}+f^{(1)})\;.
\end{equation}
The leading term in the integral above is
\begin{equation}
\label{bulk}
V^i = \frac{1}{n^{(0)}}\int d^3p\left(\frac{p}{E}\hat n^i\right)f^{(1)}\;,
\end{equation}
which is a pure first-order quantity and whose existence depends only on the 
fluctuation $f^{(1)}$. This is due to the fact that the integration over the term 
proportional to $(1-\Phi+\Psi)(p/E)\hat n^if^{(0)}$ vanishes because of isotropy of the 
unperturbed distribution function $f^{(0)}$ in momentum space.

Integration of eq.~\eqref{vlasov1-1} in momentum space is straightforward and results in
\begin{equation}
\label{momentum0}
\frac{\partial n^{(1)}}{\partial t}+\frac{1}{a}\frac{\partial(n^{(0)}V^i)}{\partial x^i}+3Hn^{(1)}+
3n^{(0)}\frac{\partial\Phi}{\partial t} = 0\;.
\end{equation}
We introduce now the particle density contrast $\delta = n^{(1)}/n^{(0)}$, which for non-relativistic particles coincides with the usual density contrast $\delta = \rho^{(1)}/\rho^{(0)}$. Using 
eq.~\eqref{continuity} governing the evolution of $n^{(0)}$, eq.~\eqref{momentum0} can be rewritten as
\begin{equation}
\dot\delta + \frac{1}{a}\partial_iV^i + 3\dot\Phi = 0\;,
\label{contrast}
\end{equation}
where $\partial_i \equiv \partial/\partial x^i$ and again overdots represent derivatives with respect to the cosmic time.

Next multiply eq.~\eqref{vlasov1-1} by $(p/E)\hat n^j$ and integrate over the momentum space. The integration
of the different terms is now more delicate and we consider it in some detail. The first term is
\begin{equation}
\frac{\partial}{\partial t}\int d^3pf^{(1)}\frac{p\hat n^j}{E}=\frac{\partial(n^{(0)}V^j)}{\partial t}\;,
\end{equation}
where we used eq.~\eqref{bulk} defining the bulk velocity. The second term is quite similar to the relation
defining the tensor $\omega^{ij}$ [see eq.~\eqref{tensoromega}] except for the fact that in 
this case the average is performed over the perturbed distribution function. Let us, for the 
moment, consider that this term defines the tensor $\omega_1^{ij}$:
\begin{equation}
\label{tensoromega1}
\frac{1}{a}\partial_i\int d^3p\frac{p^2}{E^2}\hat n^i\hat n^jf^{(1)}=\frac{1}{a}\partial_i(n^{(0)}\omega_1^{ij})\;.
\end{equation}
The third term, using a first order expansion for the energy becomes
\begin{equation}
-H\int d^3p\frac{p^2}{E}\hat n^j\frac{\partial f^{(1)}}{\partial p}
\simeq \nonumber\\ -H\int d^3p\left[\frac{p^2}{m}-\frac{p^4}{2m^3}\right]\hat n^j
\frac{\partial f^{(1)}}{\partial p}\;,
\end{equation}
Integrating by parts and keeping only the leading term $p^2/m$ gives
\begin{equation}
4H\int d^3p\frac{p\hat n^j}{m}f^{(1)} \simeq 4Hn^{(0)}V^j\;.
\end{equation}
The fourth term vanishes due to isotropy while from the fifth one it results
\begin{equation}
-\frac{1}{a}\partial_i\Psi\int d^3p\frac{\partial f^{(0)}}{\partial p}p\hat n^i\hat n^j=
\frac{n^{(0)}\delta^{ij}}{a}\partial_i\Psi\;.
\end{equation}
Collecting all these terms, the equation for the first momentum of eq.~\eqref{vlasov1-1} is
\begin{equation}
\partial_t(n^{(0)}V^j)+\frac{1}{a}\partial_i(n^{(0)}\omega_1^{ij})+4Hn^{(0)}V^j+\frac{1}{a}n^{(0)}\partial_j\Psi=0\;.
\end{equation}
Finally, combining this equation with the continuity equation for $n^{(0)}$ one obtains
\begin{equation}
\partial_tV^i + HV^i + \frac{1}{a}\partial_j\omega_1^{ij} + \frac{1}{a}\partial_i\Psi = 0\;.
\label{momentum1}
\end{equation}
An equation for the evolution of the density contrast can be derived by taking the divergence of
eq.~\eqref{momentum1} and combining it with eq.~\eqref{contrast}. After some algebra, it results in
\begin{equation}
\ddot\delta + 2H\dot\delta + 6H\dot\Phi + 3\ddot\Phi - \frac{1}{a^2}\nabla^2\Psi + 
\frac{1}{a^2}\partial_i\partial_j\omega_1^{ij} = 0\;.
\label{contrast2}
\end{equation}
The equation above is the general relativistic counterpart of that commonly found in the literature based on the Vlasov-Poisson system in a expanding background (see e.g. ref. \cite{peebles1993principles}).

\subsection{The linearised Einstein's equations}

The behaviour of the potentials $\Phi$ and $\Psi$ appearing in the perturbed metric and in the 
precedent equations is governed by Einstein's equations. The discussion of the linear Einstein's equations 
corresponding to the metric defined by eq.~\eqref{metricperturbed} can be found in different papers 
and textbooks, see for example refs.~\cite{Dodelson:2003ft,Mukhanov:2005sc}. Here we quote only the main 
results, following the approach of ref.~\cite{amendola2010dark}.

Considering a dark matter-dominated Universe, the $0-0$, the $0-i$ and the $i-i$ components 
of Einstein's equations give respectively
\begin{equation}
\label{linear00}
3H^2\Psi - 3H\dot\Phi + \frac{1}{a^2}\nabla^2\Phi = -4\pi G\rho_{\rm dm}\delta\;,
\end{equation}
\begin{equation}
\label{linear0i}
\nabla^2\left(\dot\Phi - H\Psi\right) = 4\pi G a\rho_{\rm dm}\partial_iV^i\;,
\end{equation}
and
\begin{equation}
\label{linearii}
\ddot\Phi + 3H\dot\Phi - H\dot\Psi - \left(3H^2 + 2\dot{H}\right)\Psi = 4\pi G(\delta p)\;.
\end{equation}
In these equations, the background matter density $\rho_{\rm dm} \equiv \rho^{(0)}$ and the density 
contrast $\delta$ for the particle density coincides with the mass density contrast, since we are 
considering non-relativistic matter; $(\delta p)$ is the pressure perturbation, generally written in 
terms of density contrast if an equation of state is available. From the $i-j$ component, we have an 
additional condition: $\Phi = -\Psi$, due to the absence of anisotropic stresses for DM. 

\subsection{Fourier modes}

In the next step the linear equations are written in the Fourier space. Thus, using the 
condition $\Phi = -\Psi$, eqs.~\eqref{linear00}, \eqref{linear0i} and \eqref{linearii} can be recast as
\begin{equation}
\label{fourier00}
3H^2\Psi + 3H\dot\Psi + \frac{k^2}{a^2}\Psi = -4\pi G\rho_{\rm dm}\delta\;,
\end{equation}
\begin{equation}
\label{fourier0i}
k^2\left(\dot\Psi + H\Psi\right) = 4\pi G a\rho_{\rm dm}(ik_jV^j)\;,
\end{equation}
and
\begin{equation}
\label{fourierii}
\ddot\Psi + 4H\dot\Psi + \left(3H^2 + 2\dot{H}\right)\Psi = -4\pi G(\delta p)\;,
\end{equation}
where, as usual, $k$ is the wavenumber. In order to simplify the notation, we have kept the same 
symbol for the perturbed quantities and their Fourier transforms.

Since we are considering collisionless matter, let us neglect initially the pressure and variations of pressure. Then,
from eq.~\eqref{fourierii}, combined with the background solution $a \propto t^{2/3}$ of the Einstein-de Sitter universe, one has
\begin{equation}
\ddot\Psi + 4H\dot\Psi \simeq 0\;.
\end{equation}
A solution of this equation is $\Psi (=-\Phi) =$ constant. For perturbations with scales larger than the horizon
($k \ll Ha$), eq.~\eqref{fourier00} combined with the Hubble equation leads to the well 
known result $\delta \simeq 2\Phi = -2\Psi$, i.e. the density contrast for these modes remains practically constant.

\begin{figure}[htbp]
\centering
\includegraphics[width=0.7\columnwidth]{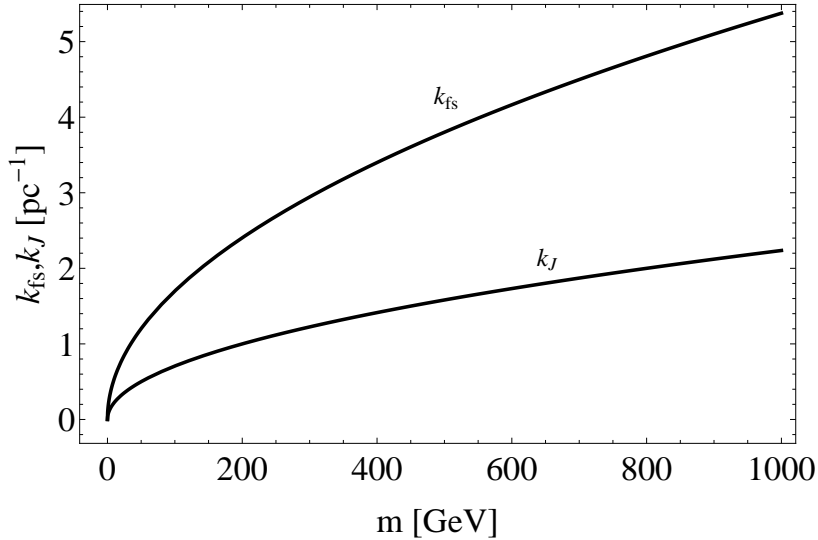}
\caption{Comparison between the comoving Jeans wavenumber derived in the present work and the comoving free streaming
wavenumber (in pc$^{-1}$) as a function of the mass of dark matter particles (in GeV). Both quantities are computed
at matter-radiation equality.}
\label{fig1}
\end{figure}

Modes corresponding to subhorizon scales ($k \gg Ha$) require the use of eq.~\eqref{contrast2}. In order 
to write this equation in the Fourier space, it is necessary to develop the term $\partial_i\partial_j\omega_1^{ik}$. As before, we assume a shear-free velocity field such as $\omega_1^{ij} = v^2_1\delta^{ij}$. Here we must be careful because although in the integral defining the tensor $\omega_1^{ij}$ [eq.~\eqref{tensoromega1}] appears only the
first order distribution function $f^{(1)}$, the normalization has to be performed with respect to
$f = f^{(0)}+f^{(1)}$ as in eq.~\eqref{bulk0}. 
In more detail, the total (i.e. unperturbed plus perturbed) velocity dispersion has, by definition, the following form:
\begin{equation}
 \sigma^2 = \frac{\int d^3p \frac{p^2}{E^2}\left(f^{(0)} + f^{(1)}\right)}{\int d^3p\left(f^{(0)} + f^{(1)}\right)} = \frac{\int d^3p \frac{p^2}{E^2}\left(f^{(0)} + f^{(1)}\right)}{n^{(0)} + n^{(1)}}\;.
\end{equation}
Developing the denominator up to first order, the above equation can be rewritten as
\begin{equation}\label{gensigmeq}
 \sigma^2 = \sigma_{(0)}^2 - \sigma_{(0)}^2\delta + v^2_1\;,
\end{equation}
where we used the definition in eq.~\eqref{tensoromega1}, together with isotropy, i.e. $\omega_1^{ij} = v^2_1\delta^{ij}$. From eq.~\eqref{gensigmeq} we can read off the first-order contribution to the velocity dispersion and write it explicitly
\begin{equation}\label{v1sigrel}
 v^2_1 = \sigma_{(0)}^2\delta + \sigma_{(1)}^2\;.
\end{equation}
The first order term $\sigma_{(1)}^2$ can be computed by perturbing eq.~\eqref{Qind}, which defines $Q$, i.e.
\begin{equation}\label{sig1rel}
 \sigma_{(1)}^2 = \frac{2}{3}\sigma_{(0)}^2\delta - \frac{2}{3}\sigma_{(0)}^2\frac{Q^{(1)}}{Q^{(0)}}\;.
\end{equation}
Since in the linear regime extreme relaxation processes like violent relaxation are expected not to take place, it is assumed as a working hypothesis that $Q$ remains constant. This is equivalent to say that the entropy per particle remains constant, cf. eq.~\eqref{Eq:Sackur-Tetrode}, what should be expected for a collisionless fluid in absence of collective dissipative mechanisms. Equations~\eqref{v1sigrel} and \eqref{sig1rel}, together with the assumption $Q^{(1)} = 0$, implies that $\sigma^2_{(1)} = (2/3)\sigma_{(0)}^2\delta$ and, finally, $v^2_1 = (5/3)\sigma_{(0)}^2\delta$.
Using this result, eq.~\eqref{contrast2} in the Fourier 
space can be written as follows:
\begin{equation}
\label{fouriercontrast}
\ddot\delta + 2H\dot\delta + \frac{k^2}{a^2}\left(\Psi - \frac{5}{3}\sigma_{(0)}^2\delta\right) = 0\;.
\end{equation} 
In the limit $k >> Ha$, eq.~\eqref{fourier00} becomes essentially the Poisson equation which, replaced 
into eq.~\eqref{fouriercontrast}, gives
\begin{equation}
\label{fouriercontrast1}
\ddot\delta + 2H\dot\delta - \left(4\pi G\rho_{\rm dm} - \frac{5}{3}\frac{k^2}{a^2}\sigma_{(0)}^2\right)\delta = 0\;.
\end{equation}
Notice that now an extra-term due to the velocity dispersion appears in what would otherwise be the 
usual equation for the evolution of perturbations of a ``pressureless'' fluid. The gravitational 
instability condition requires that the coefficient in parenthesis of eq.~\eqref{fouriercontrast1} be positive.
Thus, the critical comoving wavenumber below which perturbations grow is given by
\begin{equation}
k^2_{\rm J} = \frac{12\pi}{5}G a^2\rho_{\rm dm}\sigma^{-2}_{(0)}\;.
\label{criticalk}
\end{equation}
Neglecting numerical factors, this expression compares with that found by ref.~\cite{Boyanovsky:2008he} for
the critical free streaming wavenumber $k_{\rm fs}$. However, the physical meaning of eq.~\eqref{criticalk} is rather
different. To see this, use the definition of $Q$ to substitute the velocity dispersion in eq.~\eqref{criticalk} and
rewrite in terms of the physical critical wavelength (or physical equivalent Jeans length), i.e.,
\begin{equation}
\lambda^2_{\rm J} = \frac{5\pi}{G}\rho_{\rm dm}^{-1/3}Q^{-2/3}\;.
\label{criticaljeans}
\end{equation}
For a DM particle of $m = 200$ GeV, as can be also appreciated from fig.~\ref{fig1}, the comoving Jeans length is of the order of the parsec. Suppose now a spherical homogeneous perturbation of radius $R$ and density $\rho$. The variation of the 
gravitational potential due to a small contraction is $\Delta W \sim G\rho R^2\delta$ where again 
$\delta$ = $\Delta\rho/\rho$ is the density contrast. In the one hand, the variation of the potential energy goes into 
bulk and internal motions. Thus, the expected variation of the velocity dispersion due to such a contraction is 
$\Delta\sigma^2 \propto \Delta W \sim G\rho R^2\delta$. On the other hand, if during such a small contraction $Q$
remains constant (isentropic process) then $\Delta\sigma^2 \propto \rho^{2/3}Q^{-2/3}\delta$. For the contraction
be possible is necessary that enough gravitational energy be furnished to internal motions in order to 
satisfy the last condition. Hence, it is required that $R^2 > \rho^{-1/3}Q^{-2/3}/G$, which is essentially the
condition expressed by eq.~\eqref{criticaljeans}.

In fig.~\ref{fig1} the comoving critical wavenumber $k_{\rm J}$ derived in this work is compared with the comoving free streaming
wavenumber $k_{\rm fs}$ for different masses of DM particles. The main difference between free-streaming and the effects of a velocity dispersion resides in the physics. Free-streaming is essentially a kinematic effect: when a particle is sufficiently fast simply it escapes from the gravitational collapse. In other words, when the free-streaming time is smaller than the free-fall one, the collapse is damped. Velocity dispersion, as we show for example in eq.~\eqref{energy3}, acts as a pressure, so its physics is like the one in the Jeans mechanism, i.e. sound waves thwart the collapse.

In fig.~\ref{fig1} both quantities are computed at matter-radiation equality, when $k_{\rm fs}$ becomes practically constant while $k_{\rm J}$ increases as $a^{1/2}$. In our computations the kinetic decoupling temperature as a function of the chemical decoupling temperature and the expression for $k_{\rm fs}$ were taken from ref.~\cite{Green:2005fa}. Notice that although both scales do not differ drastically in the considered mass interval, $k_{\rm J}$ is always less than $k_{\rm fs}$, indicating that at small scales the Jeans length also controls the mass of the lowest halos. In other words, perturbations that do not satisfy the Jeans criterion do not grow and are damped by the free streaming mechanism. The Jeans mass scale is given by
\begin{equation}
M_{\rm J} = \frac{\pi^3}{2}\left(\frac{H_0^2\Omega_{\rm dm}}{k_{\rm J}^3G}\right)\;,
\end{equation}
and for a DM particle of $m = 200$ GeV the Jeans mass is about $4.3\times 10^{-6}~M_\odot$, which is of the
same order of magnitude than the mass scale derived from free streaming \cite{Green:2005fa}. The Jeans mass for
collisionless particles scales as $(1+z)^{3/2}$ as for baryons after decoupling. At thermal decoupling,
the Jeans mass for DM particles is about $10^6~M_\odot$. The derived scale is certainly relevant for future high resolution cosmological simulations since it fixes the masses of the lowest structures and substructures, which play an important role on the relaxation of halos, see e.g. refs.~\cite{Ma:2003cq,Bagla:2004qx,2012MNRAS.427L..30P}.


\section{Conclusions}\label{Sec:Concl}

In this work we discussed solutions of Vlasov-Einstein equation for a flat Friedmann background, supposed
to describe the evolution of Dark Matter collisionless particles. It is shown that after decoupling from 
the primordial plasma, the phase-space indicator $Q = \rho/(\sigma_{\rm 1D}^2)^{3/2}$ remains constant 
during the expansion of the universe (assuming non-relativistic velocities) prior to structure formation. Once 
structures begin to form, mixing processes induced by violent relaxation produce a dramatic reduction of the 
original phase-space density \cite{tremaine1986,tremaine1987erratum}. The observed value of $Q$ in the central 
regions of galaxies or clusters is only a lower limit of the primordial value established at kinetic decoupling, imposing 
only a lower bound to the mass of DM particles.

First order corrections to the energy density of DM particles, due to their velocity dispersions, include 
a kinetic term, which scales as $a^{-5}$. The maximum contribution of the latter occurs at kinetic decoupling and it 
amounts to a factor of about $10^{-4}$ of the rest energy density. In the energy conservation equation, such first order 
correction appears as an ``effective pressure'', which is equal to two thirds of the kinetic energy density, as 
for an usual Boltzmann gas.

Finally, we computed momenta of the linearised Vlasov-Einstein equation, for a perturbed flat 
Friedmann spacetime, improving past studies based on the Vlasov-Poisson system. The resulting equation for the density contrast evolution, cf. eq.~\eqref{fouriercontrast}, includes an extra term depending on the
velocity dispersion of DM particles. This correction was obtained under the assumption that the phase-space
indicator $Q$ remains constant in the linear regime, which is equivalent to say that the entropy per particle
is conserved during the growth of the density perturbations, cf. eq.~\eqref{Eq:Sackur-Tetrode}. This term permits to define the physical Jeans length for collisionless matter as a function of the primordial phase-space 
indicator $Q$, i.e $\lambda_{\rm J} = (5\pi/G)^{1/2}Q^{-1/3}\rho_{\rm dm}^{-1/6}$, above which density fluctuations are
gravitationally unstable. This length defines the minimum scale of growing perturbations above which the
gravitational energy is able to compensate the increase of the velocity dispersion of random motions due
to the adiabaticity of the process. The comoving Jeans wavenumber $k_{\rm J}$ at matter-radiation equality is about
a factor of 2-3 smaller than the wavenumber $k_{\rm fs}$ due to free streaming, controlling the cut-off scale of
the fluctuation power spectrum. If DM particles have a typical mass of 200 GeV \cite{Han:2013gba}, the associated 
Jeans mass is about $4.3\times 10^{-6}~M_\odot$, comparable with mass scales derived from free-streaming
\cite{Green:2005fa}.

\acknowledgments

OFP, DCR and JCF thank FAPES and CNPq for partial financial support. JAFP acknowledges the 
program ``Science without Borders" (CNPq) that has supported his visit to the University of Esp\'irito Santo (Brazil), where 
this project was developed.

\bibliographystyle{JHEP}
\bibliography{Vlasovbib.bib}

\providecommand{\href}[2]{#2}\begingroup\raggedright\begin{thebibliography}{10}

\bibitem{Bertone:2004pz}
G.~Bertone, D.~Hooper, and J.~Silk, {\it {Particle dark matter: Evidence,
  candidates and constraints}},  {\em Phys.Rept.} {\bf 405} (2005) 279--390,
  [\href{http://xxx.lanl.gov/abs/hep-ph/0404175}{{\tt hep-ph/0404175}}].

\bibitem{Aaij:2012ac}
{\bf LHCb Collaboration} Collaboration, R.~Aaij et~al., {\it {Strong
  constraints on the rare decays $B_s \to \mu^+ \mu^-$ and $B^0 \to \mu^+
  \mu^-$}},  {\em Phys.Rev.Lett.} {\bf 108} (2012) 231801,
  [\href{http://xxx.lanl.gov/abs/1203.4493}{{\tt arXiv:1203.4493}}].

\bibitem{Bechtle:2012zk}
P.~Bechtle, T.~Bringmann, K.~Desch, H.~Dreiner, M.~Hamer, et~al., {\it
  {Constrained Supersymmetry after two years of LHC data: a global view with
  Fittino}},  {\em JHEP} {\bf 1206} (2012) 098,
  [\href{http://xxx.lanl.gov/abs/1204.4199}{{\tt arXiv:1204.4199}}].

\bibitem{Bernabei:2010ke}
R.~Bernabei, P.~Belli, F.~Cappella, R.~Cerulli, C.~Dai, et~al., {\it {Particle
  Dark Matter in DAMA/LIBRA}},  \href{http://xxx.lanl.gov/abs/1007.0595}{{\tt
  arXiv:1007.0595}}.

\bibitem{Kelso:2011gd}
C.~Kelso, D.~Hooper, and M.~R. Buckley, {\it {Toward A Consistent Picture For
  CRESST, CoGeNT and DAMA}},  {\em Phys.Rev.} {\bf D85} (2012) 043515,
  [\href{http://xxx.lanl.gov/abs/1110.5338}{{\tt arXiv:1110.5338}}].

\bibitem{Arina:2011zh}
C.~Arina, J.~Hamann, R.~Trotta, and Y.~Y. Wong, {\it {Evidence for dark matter
  modulation in CoGeNT}},  {\em JCAP} {\bf 1203} (2012) 008,
  [\href{http://xxx.lanl.gov/abs/1111.3238}{{\tt arXiv:1111.3238}}].

\bibitem{Hooper:2012ft}
D.~Hooper, {\it {The Empirical Case For 10 GeV Dark Matter}},  {\em Phys.Dark
  Univ.} {\bf 1} (2012) 1--23, [\href{http://xxx.lanl.gov/abs/1201.1303}{{\tt
  arXiv:1201.1303}}].

\bibitem{Aprile:2010um}
{\bf XENON100 Collaboration} Collaboration, E.~Aprile et~al., {\it {First Dark
  Matter Results from the XENON100 Experiment}},  {\em Phys.Rev.Lett.} {\bf
  105} (2010) 131302, [\href{http://xxx.lanl.gov/abs/1005.0380}{{\tt
  arXiv:1005.0380}}].

\bibitem{Perivolaropoulos:2008ud}
L.~Perivolaropoulos, {\it {Six Puzzles for LCDM Cosmology}},
  \href{http://xxx.lanl.gov/abs/0811.4684}{{\tt arXiv:0811.4684}}.

\bibitem{Donato:2009ab}
F.~Donato, G.~Gentile, P.~Salucci, C.~F. Martins, M.~Wilkinson, et~al., {\it {A
  constant dark matter halo surface density in galaxies}},  {\em
  Mon.Not.Roy.Astron.Soc.} {\bf 397} (2009) 1169--1176,
  [\href{http://xxx.lanl.gov/abs/0904.4054}{{\tt arXiv:0904.4054}}].

\bibitem{DelPopolo:2010rj}
A.~Del~Popolo, {\it {On the universality of density profiles}},  {\em
  Mon.Not.Roy.Astron.Soc.} {\bf 408} (2010) 1808,
  [\href{http://xxx.lanl.gov/abs/1012.4322}{{\tt arXiv:1012.4322}}].

\bibitem{deNaray:2010zw}
R.~K. de~Naray and T.~Kaufmann, {\it {Recovering cores and cusps in dark matter
  haloes using mock velocity field observations}},  {\em
  Mon.Not.Roy.Astron.Soc.} {\bf 414} (2011) 3617,
  [\href{http://xxx.lanl.gov/abs/1012.3471}{{\tt arXiv:1012.3471}}].

\bibitem{Ogiya:2011ta}
G.~Ogiya and M.~Mori, {\it {The core-cusp problem in cold dark matter halos and
  supernova feedback: Effects of Mass Loss}},
  \href{http://xxx.lanl.gov/abs/1106.2864}{{\tt arXiv:1106.2864}}.

\bibitem{Blumenthal:1984bp}
G.~R. Blumenthal, S.~Faber, J.~R. Primack, and M.~J. Rees, {\it {Formation of
  Galaxies and Large Scale Structure with Cold Dark Matter}},  {\em Nature}
  {\bf 311} (1984) 517--525.

\bibitem{Davis:1985rj}
M.~Davis, G.~Efstathiou, C.~S. Frenk, and S.~D. White, {\it {The Evolution of
  Large Scale Structure in a Universe Dominated by Cold Dark Matter}},  {\em
  Astrophys.J.} {\bf 292} (1985) 371--394.

\bibitem{Navarro:1995iw}
J.~F. Navarro, C.~S. Frenk, and S.~D. White, {\it {The Structure of cold dark
  matter halos}},  {\em Astrophys.J.} {\bf 462} (1996) 563--575,
  [\href{http://xxx.lanl.gov/abs/astro-ph/9508025}{{\tt astro-ph/9508025}}].

\bibitem{tremaine1986}
S.~Tremaine, M.~Henon, and D.~Lynden-Bell, {\it H-functions and mixing in
  violent relaxation},  {\em Monthly Notices of the Royal Astronomical Society}
  {\bf 219} (1986), no.~2 285--297.

\bibitem{tremaine1987erratum}
S.~Tremaine, M.~H{\'e}non, and D.~Lynden-Bell, {\it Erratum: H-functions and
  mixing in violent relaxation},  {\em Monthly Notices of the Royal
  Astronomical Society} {\bf 227} (1987) 543.

\bibitem{Hogan:2000bv}
C.~J. Hogan and J.~J. Dalcanton, {\it {New dark matter physics: clues from halo
  structure}},  {\em Phys.Rev.} {\bf D62} (2000) 063511,
  [\href{http://xxx.lanl.gov/abs/astro-ph/0002330}{{\tt astro-ph/0002330}}].

\bibitem{Shao:2012cg}
S.~Shao, L.~Gao, T.~Theuns, and C.~S. Frenk, {\it {The phase space density of
  fermionic dark matter haloes}},
  \href{http://xxx.lanl.gov/abs/1209.5563}{{\tt arXiv:1209.5563}}.

\bibitem{binney1987galactic}
J.~J. Binney, {\em Galactic dynamics}.
\newblock Princeton university press, 1987.

\bibitem{Peirani:2005kw}
S.~Peirani, F.~Durier, and J.~A. De~Freitas~Pacheco, {\it {Evolution of the
  phase-space density of dark matter halos and mixing effects in merger
  events}},  {\em Mon.Not.Roy.Astron.Soc.} {\bf 367} (2006) 1011--1016,
  [\href{http://xxx.lanl.gov/abs/astro-ph/0512482}{{\tt astro-ph/0512482}}].

\bibitem{Vass:2008re}
I.~Vass, M.~Valluri, A.~Kravtsov, and S.~Kazantzidis, {\it {Evolution of the
  Dark Matter Phase-Space Density Distributions of LCDM Halos}},  {\em
  Mon.Not.Roy.Astron.Soc.} {\bf 395} (2009) 1225--1236,
  [\href{http://xxx.lanl.gov/abs/0810.0277}{{\tt arXiv:0810.0277}}].

\bibitem{Bringmann:2006mu}
T.~Bringmann and S.~Hofmann, {\it {Thermal decoupling of WIMPs from first
  principles}},  {\em JCAP} {\bf 0407} (2007) 016,
  [\href{http://xxx.lanl.gov/abs/hep-ph/0612238}{{\tt hep-ph/0612238}}].

\bibitem{Peirani:2008bu}
S.~Peirani and J.~de~Freitas~Pacheco, {\it {Dark Matter Accretion into
  Supermassive Black Holes}},  {\em Phys.Rev.} {\bf D77} (2008) 064023,
  [\href{http://xxx.lanl.gov/abs/0802.2041}{{\tt arXiv:0802.2041}}].

\bibitem{lee2010classical}
H.~Lee, {\it Classical solutions to the vlasov--poisson system in an
  accelerating cosmological setting},  {\em Journal of Differential Equations}
  {\bf 249} (2010), no.~5 1111--1130.

\bibitem{rein1997nonlinear}
G.~Rein, {\it Nonlinear stability of homogeneous models in newtonian
  cosmology},  {\em Archive for Rational Mechanics and Analysis} {\bf 140}
  (1997), no.~4 335--351.

\bibitem{Boyanovsky:2008he}
D.~Boyanovsky, H.~de~Vega, and N.~Sanchez, {\it {The dark matter transfer
  function: free streaming, particle statistics and memory of gravitational
  clustering}},  {\em Phys.Rev.} {\bf D78} (2008) 063546,
  [\href{http://xxx.lanl.gov/abs/0807.0622}{{\tt arXiv:0807.0622}}].

\bibitem{Pavlov:2012zz}
V.~Pavlov and E.~Tito, {\it {Hydrodynamical instability of dark matter:
  Analytical solution for the flat expanding universe}},  {\em Phys.Rev.} {\bf
  D85} (2012) 103010.

\bibitem{Rendall:1996gx}
A.~D. Rendall, {\it {An Introduction to the Einstein-Vlasov system}},
  \href{http://xxx.lanl.gov/abs/gr-qc/9604001}{{\tt gr-qc/9604001}}.

\bibitem{bernstein2004kinetic}
J.~Bernstein, {\em Kinetic theory in the expanding universe}.
\newblock Cambridge University Press, 2004.

\bibitem{Andreasson:2005qy}
H.~Andreasson, {\it {The Einstein-Vlasov system / kinetic theory}},  {\em
  Living Rev.Rel.} {\bf 8} (2005) 2,
  [\href{http://xxx.lanl.gov/abs/gr-qc/0502091}{{\tt gr-qc/0502091}}].

\bibitem{Okabe:2011nt}
T.~Okabe, P.~Morrison, J.~Friedrichsen, and L.~Shepley, {\it {Hamiltonian
  Dynamics of Spatially-Homogeneous Vlasov-Einstein Systems}},  {\em Phys.Rev.}
  {\bf D84} (2011) 024011, [\href{http://xxx.lanl.gov/abs/1106.4807}{{\tt
  arXiv:1106.4807}}].

\bibitem{peebles1993principles}
P.~J.~E. Peebles, {\em Principles of physical cosmology}.
\newblock Princeton University Press, 1993.

\bibitem{Boyanovsky:2007ay}
D.~Boyanovsky, H.~de~Vega, and N.~Sanchez, {\it {Constraints on dark matter
  particles from theory, galaxy observations and N-body simulations}},  {\em
  Phys.Rev.} {\bf D77} (2008) 043518,
  [\href{http://xxx.lanl.gov/abs/0710.5180}{{\tt arXiv:0710.5180}}].

\bibitem{Dalcanton:2000hn}
J.~J. Dalcanton and C.~J. Hogan, {\it {Halo cores and phase space densities:
  Observational constraints on dark matter physics and structure formation}},
  {\em Astrophys.J.} {\bf 561} (2001) 35--45,
  [\href{http://xxx.lanl.gov/abs/astro-ph/0004381}{{\tt astro-ph/0004381}}].

\bibitem{davis1977integration}
M.~Davis and P.~Peebles, {\it On the integration of the bbgky equations for the
  development of strongly nonlinear clustering in an expanding universe},  {\em
  The Astrophysical Journal Supplement Series} {\bf 34} (1977) 425--450.

\bibitem{Davis:1982gc}
M.~Davis and P.~Peebles, {\it {A Survey of galaxy redshifts. 5. The Two point
  position and velocity correlations}},  {\em Astrophys.J.} {\bf 267} (1982)
  465--482.

\bibitem{Kolb:1990vq}
E.~W. Kolb and M.~S. Turner, {\it {The Early Universe}},  {\em Front.Phys.}
  {\bf 69} (1990) 1--547.

\bibitem{Dodelson:2003ft}
S.~Dodelson, {\it {Modern cosmology}}, .

\bibitem{Mukhanov:2005sc}
V.~Mukhanov, {\it {Physical foundations of cosmology}}, .

\bibitem{Hofmann:2001bi}
S.~Hofmann, D.~J. Schwarz, and H.~Stoecker, {\it {Damping scales of neutralino
  cold dark matter}},  {\em Phys.Rev.} {\bf D64} (2001) 083507,
  [\href{http://xxx.lanl.gov/abs/astro-ph/0104173}{{\tt astro-ph/0104173}}].

\bibitem{huang1987statistical}
K.~Huang, {\it Statistical mechanics, 2nd},  {\em Edition (New York: John Wiley
  \& Sons)} (1987).

\bibitem{amendola2010dark}
L.~Amendola and S.~Tsujikawa, {\em Dark energy: theory and observations}.
\newblock Cambridge University Press, 2010.

\bibitem{Green:2005fa}
A.~M. Green, S.~Hofmann, and D.~J. Schwarz, {\it {The First wimpy halos}},
  {\em JCAP} {\bf 0508} (2005) 003,
  [\href{http://xxx.lanl.gov/abs/astro-ph/0503387}{{\tt astro-ph/0503387}}].

\bibitem{Ma:2003cq}
C.-P. Ma and E.~Bertschinger, {\it {A Cosmological kinetic theory for the
  evolution of cold dark matter halos with substructure: Quasilinear theory}},
  {\em Astrophys.J.} {\bf 612} (2004) 28--49,
  [\href{http://xxx.lanl.gov/abs/astro-ph/0311049}{{\tt astro-ph/0311049}}].

\bibitem{Bagla:2004qx}
J.~S. Bagla, J.~Prasad, and S.~Ray, {\it {Gravitational collapse in an
  expanding background and the role of substructure. 1. Planar collapse}},
  {\em Mon.Not.Roy.Astron.Soc.} {\bf 360} (2005) 194--202,
  [\href{http://xxx.lanl.gov/abs/astro-ph/0408429}{{\tt astro-ph/0408429}}].

\bibitem{2012MNRAS.427L..30P}
S.~V. {Pilipenko}, A.~G. {Doroshkevich}, V.~N. {Lukash}, and E.~V. {Mikheeva},
  {\it {Effect of small-scale density perturbations on the formation of dark
  matter halo profiles}},  {\em Mon.Not.Roy.Astron.Soc.} {\bf 427} (Nov., 2012)
  L30--L34, [\href{http://xxx.lanl.gov/abs/1209.2682}{{\tt arXiv:1209.2682}}].

\bibitem{Han:2013gba}
T.~Han, Z.~Liu, and A.~Natarajan, {\it {Dark Matter and Higgs Bosons in the
  MSSM}},  \href{http://xxx.lanl.gov/abs/1303.3040}{{\tt arXiv:1303.3040}}.

\end{thebibliography}\endgroup

\label{lastpage}

\end{document}